\documentclass[onecolumn]{pasj01}

\usepackage{siunitx}
\usepackage[nooneline]{subfigure}
\usepackage{ulem}
\usepackage{xcolor}
\subfiguretopcaptrue
\usepackage[switch,mathlines]{lineno}
\Received{}%{yyyy/mm/dd}
\Accepted{}%{yyyy/mm/dd}
\definecolor{royalpurple}{rgb}{0.47, 0.32, 0.66}
\begin{document} 
\title{Possible X-ray Cocoon Emission from GRB~050709}
%\LETTERLABEL %%% <-- uncomment for LETTER article  
%\REVIEWLABEL %%% <-- uncomment for REVIEW article  

%%% begin:list of authors
% Do NOT capitalize all letters in "textsc".
\author{Naoki \textsc{Ogino}\altaffilmark{1}%
%\thanks{Example: Present Address is xxxxxxxxxx}
}
\altaffiltext{1}{Faculty of Mathematics and Physics, Institute of Science and Engineering, Kanazawa University, Kanazawa 920-1192, Japan}
\email{naokiogino3e8@gmail.com}

\author{Makoto \textsc{Arimoto},\altaffilmark{1,2}}
%\altaffiltext{2}{B-Address of Institute}
\email{arimoto@se.kanazawa-u.ac.jp}
\altaffiltext{2}{Advanced Research Center for Space Science and Technology, Institute of Science and Engineering, Kanazawa University, Kanazawa 920-1192, Japan}
\author{Hamid \textsc{Hamidani}\altaffilmark{3}}
\altaffiltext{3}{Astronomical Institute, Graduate School of Science, Tohoku University, Sendai 980-8578, Japan}
\author{Takanori \textsc{Sakamoto}\altaffilmark{4}}
\altaffiltext{4}{Department of Physics and Mathematics, College of Science and Engineering, Aoyama Gakuin University, Kanagawa 252-5258, Japan}

\author{Daisuke \textsc{Yonetoku}\altaffilmark{1,2}}

\author{Tatsuya \textsc{Sawano}\altaffilmark{1,2}}

\author{Motoko \textsc{Serino}\altaffilmark{4}}

\author{Katsuaki \textsc{Asano}\altaffilmark{5}}
\altaffiltext{5}{Institute for Cosmic Ray Research, The University of Tokyo, Chiba 277-8582, Japan}

\author{Nobuyuki \textsc{Kawai}\altaffilmark{6}}
\altaffiltext{6}{Graduate School of Science, Tokyo Institute of Technology, Tokyo 152-8551, Japan}

%\email{ccccc@xxx.xxx.xx.xx}
%%% end:list of authors

%% `\KeyWords{}' always has to be placed before ``\maketitle'' 
%%  List of Key Words:  https://academic.oup.com/pasj/pages/Pasj_Keywords 
\KeyWords{gamma-ray burst: indivisual (GRB~050709) --- gravitational waves --- stars: neutron}

\maketitle

%Please read ``IMPORTANT NOTICE'' carefully before preparing a manuscript. 
% \noindent IMPORTANT NOTICE\\
% 1. ``\verb|\draft|'' creates single column and double spaces format.\\
% 2. If you comment out ``\verb|\draft|'', the output will be double column
%   and single space.\\
% 3. For cross-references, the use of ``\verb|\label|, \verb|\ref|, \verb|\cite|'' 
%   and the thebibliography environment is strongly recommended. \\
% 4. Do NOT use ``\verb|\def|, \verb|\renewcommand|''.\\
% 5. Do NOT redefine commands provided by PASJ01.cls.\\

\begin{abstract}
The detection of the short gamma-ray burst (SGRB) 050709 by the \textit{HETE-2} satellite opened a new window into understanding the nature of SGRBs, offering clues about their emission mechanism and progenitors, with the crucial aid of optical follow-up observations.
Here, we revisit the prompt emission of GRB 050709.
Our analysis reveals an initial hard spike $\sim$200\,ms long, followed by a subsequent soft tail emission lasting $\sim$300\,ms. These components could be common among other SGRBs originating from binary neutron merger events, such as GW/GRB~170817A.
Detailed temporal and spectral analyses indicate that the soft tail emission might be attributed to the cocoon formed by the relativistic jet depositing energy into the surrounding material. 
We find the necessary cocoon parameters at the breakout, as consistent with numerical simulation results.
We compared the physical parameters of this cocoon with those of other SGRBs.
The relatively higher cocoon pressure and temperature in GRB~050709 may indicate a more on-axis jet compared to GRB~170817A and GRB~150101B.
\end{abstract}

%\pagewiselinenumbers

\section{Introduction}
The observation of the gravitational-wave event GW~170817 \citep{Abbott+2017a} marks a major historic event as it represents not only the first discovery of a gravitational-wave signal from a binary neutron star (BNS) merger, 
but also the detection of its electromagnetic (EM) counterparts. 
This discovery has opened a new window in multi-messenger astronomy.
Thanks to these multi-wavelength observations, 
the following EM counterparts were associated with GW/GRB~170817A:
(1) ``short gamma-ray bursts (SGRBs)'' (i.e., prompt emission) in the X-ray and ${\gamma}$-ray bands  \citep{Abbott+2017b,Goldstein+2017}, (2) ``kilonovae (macronovae)'' in the ultraviolet, visible, and infrared bands \citep{2017ApJ...848L..33A, 2017ApJ...848L..19C,Coulter+2017,2017ApJ...848L..29D,2017Sci...358.1570D, 2017Sci...358.1583K,Kasliwal+2017, 2017ApJ...848L..18N,2017Natur.551...67P,2017Natur.551...75S,2017Sci...358.1574S,2017ApJ...848L..16S,2017PASJ...69..102T,2017PASJ...69..101U,2017ApJ...848L..24V}
and (3) ``afterglows'' from the radio to X-ray bands \citep{DAvanzo+2018,Resmi+2018,Mooley+2018,Lamb+2019,2019Sci...363..968G,2019MNRAS.489.1919T}. 

For the prompt emission of GRB~170817A, $\sim$1.7\,s after the BNS merger time, gamma-rays photons were observed by \textit{Fermi}/GBM for a duration of $\sim2$\,s \citep{Goldstein+2017} and \textit{INTEGRAL}/SPI-ACS with a ${\sim}3{\sigma}$ confidence level \citep{Savchenko+2017}.
The observed isotropic equivalent luminosity is $\sim$10$^{47}$ erg/s which is 3--4 orders of magnitude lower than typical SGRBs \citep{Abbott+2017b}, while the fluence, peak flux, peak energy $E_{\rm peak}$, and duration of GRB~170817A are all roughly consistent with the standard SGRBs \citep{Goldstein+2017}.
The time-resolved analysis of the prompt emissions revealed the existence of a short hard pulse with a non-thermal spectrum followed by a soft tail with a blackbody spectrum  of $\sim2$\,s \citep{Goldstein+2017}.

In recent years, analysis of SGRB events with similar temporal and spectral properties to those of GRB~170817A has been performed \citep{Burns+2018, vonKienlin+2019}. 
\citet{Burns+2018} conducted the time-resolved analysis  for
the prompt emission of the nearby SGRB~150101B with $z=0.134$ using the \textit{Fermi}/GBM  data. 
They also found a short hard-spike component and a long soft-tail one, which is similar to GRB~170817A. 
The \textit{Fermi}/GBM spectral analysis indicates that the soft-tail emission has a thermal origin. 
In addition, a kilonova was discovered for GRB~150101B \citep{2018NatCo...9.4089T}.
Thus, these two nearby SGRBs with kilonovae have in common a soft-tail component in their prompt emission, suggesting that the soft tail could be a ubiquitous property of SGRBs produced by BNS mergers.
Note that in the early days of GRB observations, a soft tail was also detected from the April 27 burst in 1972 by the historical Apollo 16 observation \citep{1974ApJ...194L..19M}.

The \textit{High Energy Transient Explorer-2} (\textit{HETE-2}) satellite \citep{Ricker+2001} detected the short gamma-ray burst GRB~050709 \citep{Villasenor+2005}.
This observation led to the very first detection of 
an optical counterpart to a short GRB \citep{2006A&A...447L...5C}.
It was located in the outskirt of a host galaxy with an offset of 3.8 kpc from the center of the galaxy, at a redshift $z=0.16$, with non-detection of a supernova component \citep{Fox+2005}. This opened a new window to multi-wavelength observations of SGRBs.
In addition, \citet{Jin+2016} reported that the late optical emission observed at $t_{\rm obs} > 2.5$\,days after GRB~050709 (in the observer frame) can be explained by a kilonova transient,
Here, we present a reanalysis of the X-ray and $\gamma$-ray data of GRB~050709 observed by \textit{HETE-2}.
We found that it consists of a hard-spike followed by a soft-tail component in its prompt emission,  in similarity to GRB~150101B and GRB~170817A.
Thus, we focused on the temporal and spectral features of the soft-tail component, which had not been analyzed in previous studies.
Our detailed spectral analysis shows that the soft-tail component likely originates from the thermal emission caused by the jet energy injection into the expanding ejecta material  
(i.e., cocoon emission; \cite{2006ApJ...652..482P, 2017ApJ...834...28N, 2017MNRAS.471.1652L, 2017ApJ...848L...6L, 2018MNRAS.479..588G, 2018ApJ...867...18N, 2018PTEP.2018d3E02I, 2018PhRvL.120x1103L, 2018ApJ...855..103P, 2020MNRAS.491.3192H, Hamidani+2021,2024ApJ...963..137H}).
Furthermore, we estimated the physical parameters of the cocoon of GRB~050709, and compared them with those of GRB~150101B and GRB~170817A.

This paper is organized as follows: the observation of the \textit{HETE-2} and the detection of the soft-tail component are described in Section \ref{sec:observation}, the physical parameters of the cocoon emission is quantified in Section \ref{sec:orig_ST}, the origin of the extended emission is discussed in Section \ref{sec:orig_EE}, and  
our conclusion is presented in Section \ref{sec:conclusion}. 
In this paper, errors are defined as within 90\% confidence level.
%Errors correspond to the 90\% confidence region throughout this paper\hhh{??? I did not understand this last sentence, what errors? please specify more; perhaps you want to say "In this paper, errors are defined as within 90\% confidence level..." or so, please clarify it???}.

\section{Observed Properties of GRB~050709}
\label{sec:observation}

\subsection{Overview of the observations}
GRB~050709 was detected by the Wide-Field X-Ray Monitor (WXM, 2--25\,keV;  \cite{Shirasaki+2003}) and the French Gamma Telescope (FREGATE, 6--400\,keV; \cite{Atteia+2003}) on board of the \textit{HETE-2} on July 9 2005, at 22:36:37 UT (denoted as $t_{\rm 0}$) \citep{Villasenor+2005}.
The WXM location was found by ground analysis to be right ascension (RA) +23\,h\,01\,min\,44\,s and declination (dec.) $\ang{-38;59;52}$ (J2000) with an error radius of $\ang{;14.5;}$.
The X-ray afterglow was observed by \textit{Chandra} at $t_{\rm 0} + {\sim}2.5, 16.0, 16.1$\,days and \textit{Swift} at $t_{\rm 0} + {\sim}1.6, 2.4, 3.2, 4.3$\,days.
 The \textit{Swope-40} and  \textit{Subaru} telescope performed the optical observations in the $i^\prime$ and $K^\prime$ bands, respectively, ${\sim}1.5$\,days after the trigger \citep{Fox+2005}. 
 Notably,  the Gemini Multi-Object Spectrograph on the Gemini North telescope identified the redshift of the host galaxy to be $z=0.16$ \citep{Fox+2005}.
The \textit{Hubble Space Telescope} (\textit{HST}) carried out four observations  in the F814W band during the initial one month after the burst \citep{Fox+2005} and found a fading optical emission likely originating from a kilonova  \citep{Jin+2016}.
 One year after the burst, the \textit{HST} did not detect significant emission related to this GRB in the F814W band at the same position \citep{Jin+2016}. 
The \textit{Very Large Telescope} (\textit{VLT}) also detected an optical afterglow in the $V$, $R$ and $I$ bands  from July 12 to 30, 2005 \citep{2006A&A...447L...5C}.

\subsection{Prompt emission phase}
\subsubsection{Lightcurves}
In Fig. \ref{fig:lc_prompt}, we show the light curves of the prompt emission phase observed with WXM (2--25\,keV) and FREGATE (6--400\,keV).
The durations of the prompt emission in the  2--25 and 30–-400 keV energy bands are $T_{\rm 90}$ = 220$\pm$50 ms and 70$\pm$10 ms, respectively \citep{Villasenor+2005}, where $T_{90}$ is measured as the duration of the time interval during which 90\% of the total observed counts have been detected.
Our reanalysis clearly shows for the first time that the lightcurve in the initial 0.5-s interval  consists of two components in the prompt emission phase as seen in GRB~170817A and GRB~150101B: (1) ``Hard Spike (HS)'', which was observed at  $t_{\rm 0} + 0.0-0.2$\,s in 
all energy bands (2--400 keV); and (2) ``Soft Tail (ST)'', which was observed only in the soft X-ray energy range of 2--10\,keV at  $t_{\rm 0} + 0.2-0.5$\,s.

\subsubsection{Significance of the soft tail emission}
The detection significance of the soft-tail emission is briefly calculated to be ${\sim}5.6{\sigma}$ using the light curve in the 2--10\,keV band considering the X-ray photon statistics of the GRB emission and the background fluctuation\footnote{The background count rate of the light curve in the 2--10 keV band is ${\sim}$300\,counts/s. The background-subtracted count rate of the soft-tail component at $t_{\rm 0} + $ 250--400\,ms is ${\sim}$250\,counts/s. Thus, the signal to noise ($S/N$) ratio can be calculated as $S/N$ = (250\,counts/s $\times$ 0.15 s)/(300 counts/s $\times$ 0.15 s)$^{1/2}$ $\sim$ 5.6.}. 
In addition, by using the \textit{HETE-2} science tools for transient localization, we calculated the incident X and Y angles of X-rays from the hard-spike emission with respect to the WXM boresight
and found it as ${\theta_{\rm x}^{\rm HS}}=-14.27^{+0.08}_{-0.03}$\,deg and ${\theta_{\rm y}^{\rm HS}}=-2.12^{+0.03}_{-0.06}$\,deg, respectively.
The incident X and Y angles of X-rays from the soft tail emission were also calculated 
and found as $\theta_{\rm x}^{\rm ST}=-14.48^{+24.42}_{-0.14}$\,deg and ${\theta_{\rm y}^{\rm ST}}=-2.13^{+33.60}_{-28.58}$\,deg, respectively, for the soft tail. The obtained incident angles are consistent between the hard-spike and soft-tail emissions. Note that the uncertainty of the incident Y angle of the soft-tail emission is very large due to limitations of the X-ray photon statistics.
Therefore, we concluded that the soft-tail emission is most likely a genuine source signal of this GRB, rather than a statistical fluctuation.

\subsubsection{Spectra}
For the spectral analysis, we used XSPEC version 12.12.0 \citep{1996ASPC..101...17A} with the WXM and FREGATE data, and adopted the ``PGstat'' statistic, suitable for low photons counts \citep{2011hxra.book.....A}. 
We adopted four types of spectral models, i.e., power-law (PL), cutoff PL (CPL), blackbody (BB), and multi-temperature blackbody (DISKPBB). 
Hence, the DISKPBB model was originally used for reproducing emission from an accretion disk with multiple-temperature blackbody components. This emission is described by $kT_{\rm disk}(r) \propto r^{-p}$, where  $k$ is the Boltzmann constant, $T_{\rm disk}(r)$ represents the local disk temperature, $r$ is the distance from a compact object, and $p$ is the power-law index.
The temperature at the inner disk radius is denoted by $kT_{\rm in}$.
In this study, we used the DISKPBB model implemented in XSPEC to represent a modified blackbody spectrum broadened by the relativistic motion of the ejecta originating from the cocoon empirically\footnote{It is worth noting that, in this context, the cocoon may not have a single temperature distribution and may be emitting multi-temperature blackbody radiation (i.e.,  non-local thermodynamic equilibrium).}. Similar approaches were employed in previous studies \citep{2010ApJ...709L.172R, 2016ApJ...833..139A}. 

\begin{figure}
    \centering
    \includegraphics[width=0.98\textwidth]{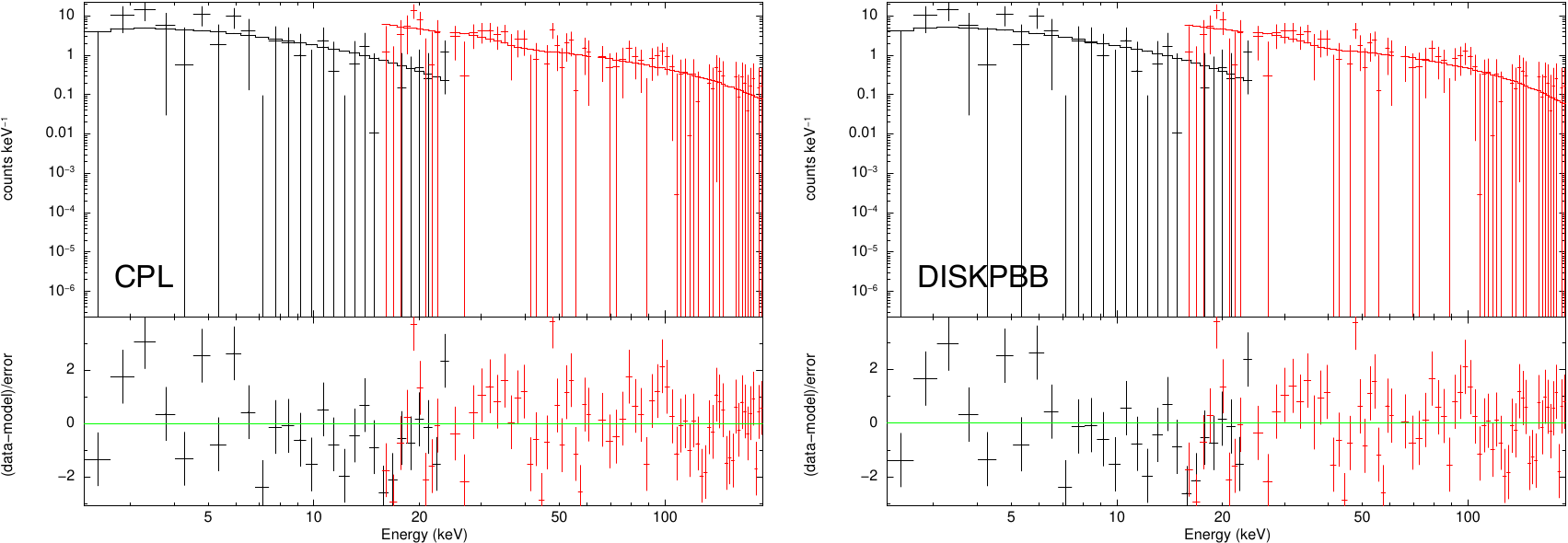}
    \caption{
    Spectral fitting and the respective residuals from $t_{\rm 0}$ +0.2 s to +0.5 s (soft tail) of GRB~050709 observed with WXM (black) and FREGATE (red). The left and right spectra are fitted by cutoff power-law function (CPL) and disk blackbody (DISKPBB), respectively.
    }
    \label{fig:soft_tail_spec}
\end{figure}

Using these functions, our fitting results are shown in Table \ref{tab:spec_prompt}.
The hard spike (HS) and the soft tail (ST) are best fit by  a cutoff power-law with $E_{\mathrm{peak}}^{\mathrm{HS}} = 200^{+437}_{-81}$\,keV and a multi-temperature blackbody with $kT_{\rm in} = 48.5^{+15.2}_{-18.5}$ keV, respectively.
%, where $k$ is the Boltzmann constant and $T_{\rm in}$ is the temperature at the inner disk radius of the DISKPBB model.
According to this best fit, the isotropic energy of the hard spike is $E_{\mathrm{iso}}^{\mathrm{HS}}=4.41^{+0.60}_{-0.32}\times10^{49}\,\mathrm{erg}$, and the isotropic luminosity of the hard spike is $L_{\mathrm{iso}}^{\mathrm{HS}}=2.56^{+0.35}_{-0.19}{\times}10^{50}\,\mathrm{erg\,s^{-1}}$.
In addition, for the soft tail emission spectrum shown in Fig. \ref{fig:soft_tail_spec}, the PGStat values of the CPL and DISKPBB models have similar values, i.e., PGstat = 93.8 (dof = 97) for CPL and PGstat = 92.7 (dof = 97) for DISKPBB. The difference of the PGstat values ($\Delta$PGstat $\sim$ 1) is not so statistically significant.
Thus, the CPL is almost equivalent to the DISKPBB model, and the effective temperature of the CPL can be found to be $kT_{\rm eff} = E_{\rm peak}/2.8 \sim 42$\,keV using Wien's law, which is consistent with the temperature obtained from the DISKPBB model ($kT_{\rm in} = 48.5^{+15.2}_{-18.5}$ keV).

\begin{figure*}
    \centering
    \includegraphics[width=0.65\textwidth]{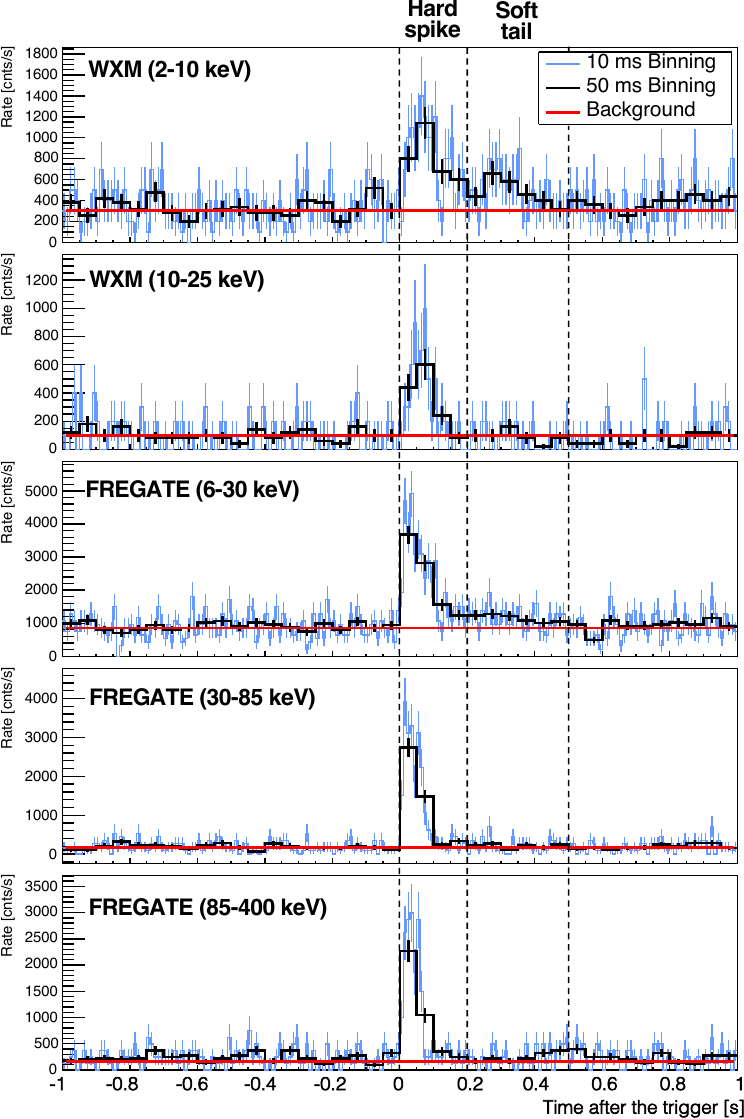} 
\caption{Composite light curve of GRB~050709 from $t_{\rm 0}$ $-1$\,s to $+1$\,s in different energy ranges with 10\,ms/bin (blue line) and 50\,ms/bin (black line), showing the short hard spike and the soft tail. The vertical lines represent the time intervals of the hard spike and the soft tail. The red line represents the background, which is the average of the count rate from $t_{\rm 0}$  - 1\,s to  -0.5\,s. We show that the soft tail is clearly detected with a significance of $\sim5.6\sigma$ 
using a binning of 50\,ms for the light curve of 2--10\,keV (WXM).} 
\label{fig:lc_prompt}
\end{figure*}

\begin{table*}
 \tbl{Results of Spectral Fitting of the Prompt Emission of GRB~050709.}{%
 \resizebox{\textwidth}{!}{
 \begin{tabular}{cccccccccc} \hline \hline
    Interval from $t_{\rm 0}$ & Energy band & Model        & $E_{\mathrm{peak}}$ & Photon index, $p$                & $kT_{\rm in}$               & PGstat/dof           & Energy flux                          & $L_\mathrm{iso}$               & $E_\mathrm{iso}$       \\
    (s)                       &  (keV)      &              & (keV)               &                                  & (keV)                       &                      & ($10^{-7}$\,\si{erg.cm^{-2}.s^{-1}}) & ($10^{49}$\,\si{erg.s^{-1}})   & ($10^{49}$\,erg)       \\ \hline
    0--0.2                    & 2--400      & CPL          & $200^{+437}_{-81}$  & $1.23^{+0.14}_{-0.15}$           &                             & 174.0/142            & $39.5^{+5.4}_{-2.9}$                 & $25.6^{+3.5}_{-1.9}$           & $4.41^{+0.60}_{-0.32}$ \\
    (Hard Spike)              &   2--400    & DISKPBB       &                     & $0.59^{+0.01}_{-0.02}$ & $141.7^{+8.3}_{-63.3}$      & 177.4/142    &                                      &                                &                         \\\hline
    0.2--0.5                  &  2--400     & CPL          & $117^{+56}_{-55}$   & $1.20^{+0.36}_{-0.52}$           &                             & 93.8/97              & $9.12^{+2.04}_{-2.14}$               & $5.90^{+1.32}_{-1.38}$         & $1.02^{+0.23}_{-0.24}$ \\
    (Soft Tail)               &    2--400   & BB           &                     &                                  & $16.5^{+3.4}_{-2.8}$        & 110.5/98             &                                      &                                &                        \\
                              &  2--400    & DISKPBB      &                     & $0.63^{+0.10}_{-0.05}$      & $48.5^{+15.2}_{-18.5}$ & 92.7/97         & $8.79^{+2.11}_{-1.93}$       & $5.69^{+1.37}_{-1.25}$ & $0.98^{+0.23}_{-0.22}$\\ 
                              &    8--400   & CPL     & $76.2^{+56.5}_{-17.8}$  & $0.01^{+0.96}_{-1.32}$            &         & 76.0/85             &                                      &                                &                        \\
                              &    8--400   & BB      &                     &                                  & $16.9^{+3.5}_{-2.8}$        & 79.4/86          &                                      &                                &                        \\
                              &  8--400   & DISKPBB &                     & $0.90^{+0.46}_{-0.46}$     & $29.9^{+6.0}_{-6.0}$ & 76.2/85         &       &  \\ 
       \hline 
 \end{tabular}}}
 \label{tab:spec_prompt}
\end{table*}

\subsection{Extended Emission}
\begin{figure*}
    \centering
    \includegraphics[width=0.65\textwidth]{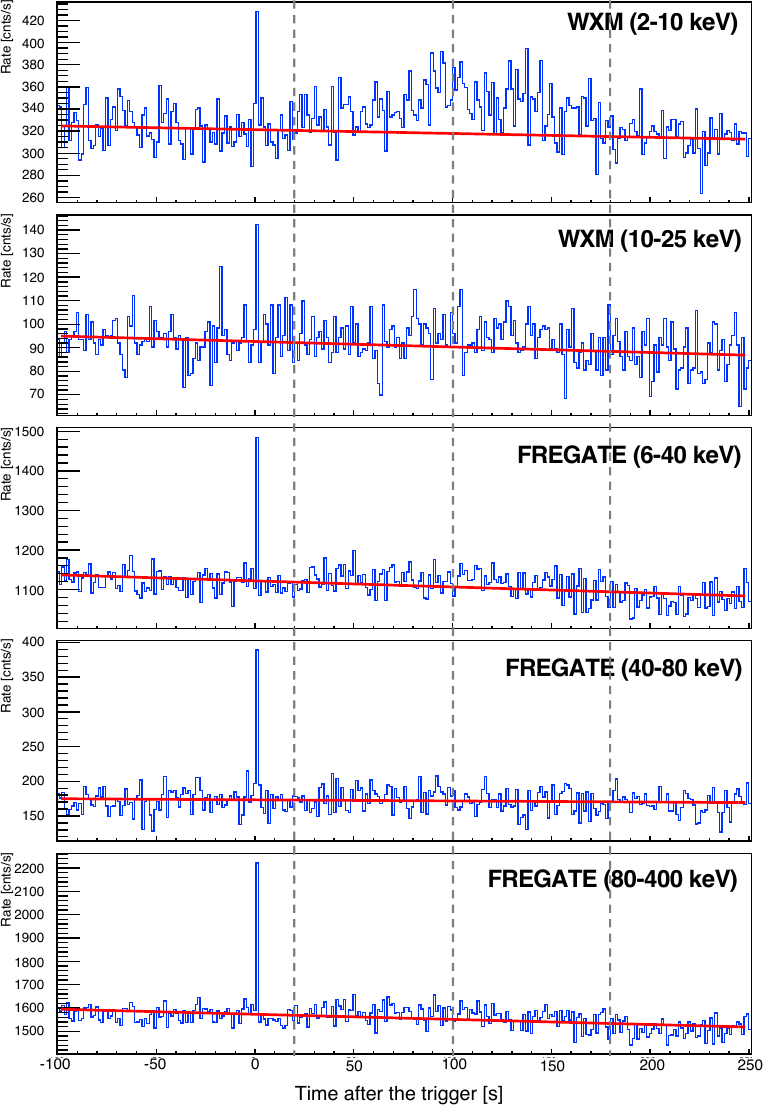}
    \caption{The composite light curve of GRB~050709 from $t_{\rm 0}$ -100\,s to $+250$\,s in the 2--400 keV band with 1-s bin (blue line). The red line represents the background model calculated by the fitting using a linear function. We see the soft X-ray emission at $t_{\rm 0}+20$--$180$\,s in the energy range of 2--10\,keV (WXM). The vertical dashed lines represent the time intervals referred to in Table \ref{tab:EE}.
    } 
    \label{fig:EE}
\end{figure*}

Some SGRBs are followed by a longer, lasting ${\sim}100$\,s , but a softer X-ray emission \citep{Norris+2006}.
The extended emission is present in GRB 050709 in the 2--10\,keV band shown in Fig. \ref{fig:EE}.
According to \cite{Villasenor+2005}, its duration is $T_{90}=130\pm7$\,s in the 2--25\,keV band.
As it can be seen in Fig. \ref{fig:EE}, the count rate of the background is 
low and fluctuating; thus the light curves were drawn for each channel and the background was subtracted by fitting with a linear function as performed in the \textit{Fermi}/GBM analysis \citep{Biltzinger+2020}.
We performed the time-resolved spectral analysis in the early ($t_{\rm 0}$ +20--100\,s) and late phases ($t_{\rm 0}$ + 100--180\,s) using the same spectral models of the prompt emission.
Table \ref{tab:EE} shows the results of the spectral analysis of the extended emission and we find that it is hard to provide useful constraints on the spectral model due to poor photon statistics. Thus, we focus on the spectral property of the PL, characterized by a time-independent photon index of $\sim$2.4. 
The obtained fitting results may indicate that there is no spectral softening for this burst, although the two time intervals of the extended emission was arbitrarily defined and the uncertainties of the fitted parameters were too large to discuss their significance.
Here,  the extended emission of many SGRBs detected by \textit{Swift} shows a significant spectral softening in previous studies (e.g., \cite{Kagawa+2015,2019ApJ...877..147K}).
Accordingly, for the extended emission of this GRB, we find the isotropic energy as $E_\mathrm{iso} = 3.36^{+0.17}_{-0.43}{\times}10^{50}$\,erg and the isotropic luminosity as $L_\mathrm{iso} = 2.44^{+0.13}_{-0.31}{\times}10^{49}$\,\si{erg.s^{-1}}. 

\begin{table*}
 \tbl{Spectral Fitting Results of Extended Emission}{%
 \begin{tabular}{cccccc} \hline \hline
    Interval from $t_{\rm 0}$       & Model        & $E_{\mathrm{peak}}$    & Photon index, $p$                       & $kT$                           & PGstat/dof \\
    (s)                             &              & (keV)                  &                                   & (keV)                          &            \\ \hline
    20--180                         & CPL          & $3.33^{+1.40}_{-1.12}$ & $1.61^{+0.36}_{-0.86}$            &                                & 39.5/49    \\
    (Whole)                         & PL           &                        & $2.43^{+0.14}_{-0.13}$            &                                & 41.4/50    \\
                                    & BB           &                        &                                   & $1.22^{+0.11}_{-0.10}$         & 44.1/50    \\
                                    & DISKPBB &                        & $0.468^{+0.031}_{-0.031}$ & $19.9^{+6.0}_{-13.8}$  & 40.9/49    \\ \hline
%                                   & DISKBB       &                        &                                   & $2.01^{+0.23}_{-0.21}$         & 41.3/50    \\ \hline
    20--100                         & CPL          & $4.94^{+1.62}_{-1.23}$ & $1.67^{+0.32}_{-0.31}$            &                                & 37.4/49    \\
    (Early phase)                   & PL           &                        & $2.35^{+0.22}_{-0.21}$            &                                & 41.1/50    \\
                                    & BB           &                        &                                   & $1.36^{+0.19}_{-0.16}$         & 48.4/50    \\
%                                   & DISKBB       &                        &                                   & $2.25^{+0.36}_{-0.40}$         & 47.6/50    \\ \hline
                                    & DISKPBB &                        & $0.481^{+0.031}_{-0.063}$ & $33.7^{+4.8}_{-29.6}$  & 41.4/49    \\ \hline
    100--180                        & CPL          & $2.71^{+2.11}_{-1.20}$ & $1.67^{+0.26}_{-0.89}$            &                                & 40.9/49    \\
    (Late phase)                    & PL           &                        & $2.44^{+0.23}_{-0.21}$            &                                & 41.3/50    \\
                                    & BB           &                        &                                   & $1.16^{+0.17}_{-0.14}$         & 45.2/50    \\
%                                   & DISKBB       &                        &                                   & $2.00^{+0.37}_{-0.39}$         & 43.2/50    \\ \hline
                                    & DISKPBB &                        & $0.465^{+0.047}_{-0.064}$ & $10.0^{+2.4}_{-6.1}$   & 41.5/49    \\ \hline
 \end{tabular}}\label{tab:EE}
\end{table*}

\subsection{Minimum Variability Timescale}
\begin{figure}
    \centering
    \includegraphics[width=0.55\textwidth]{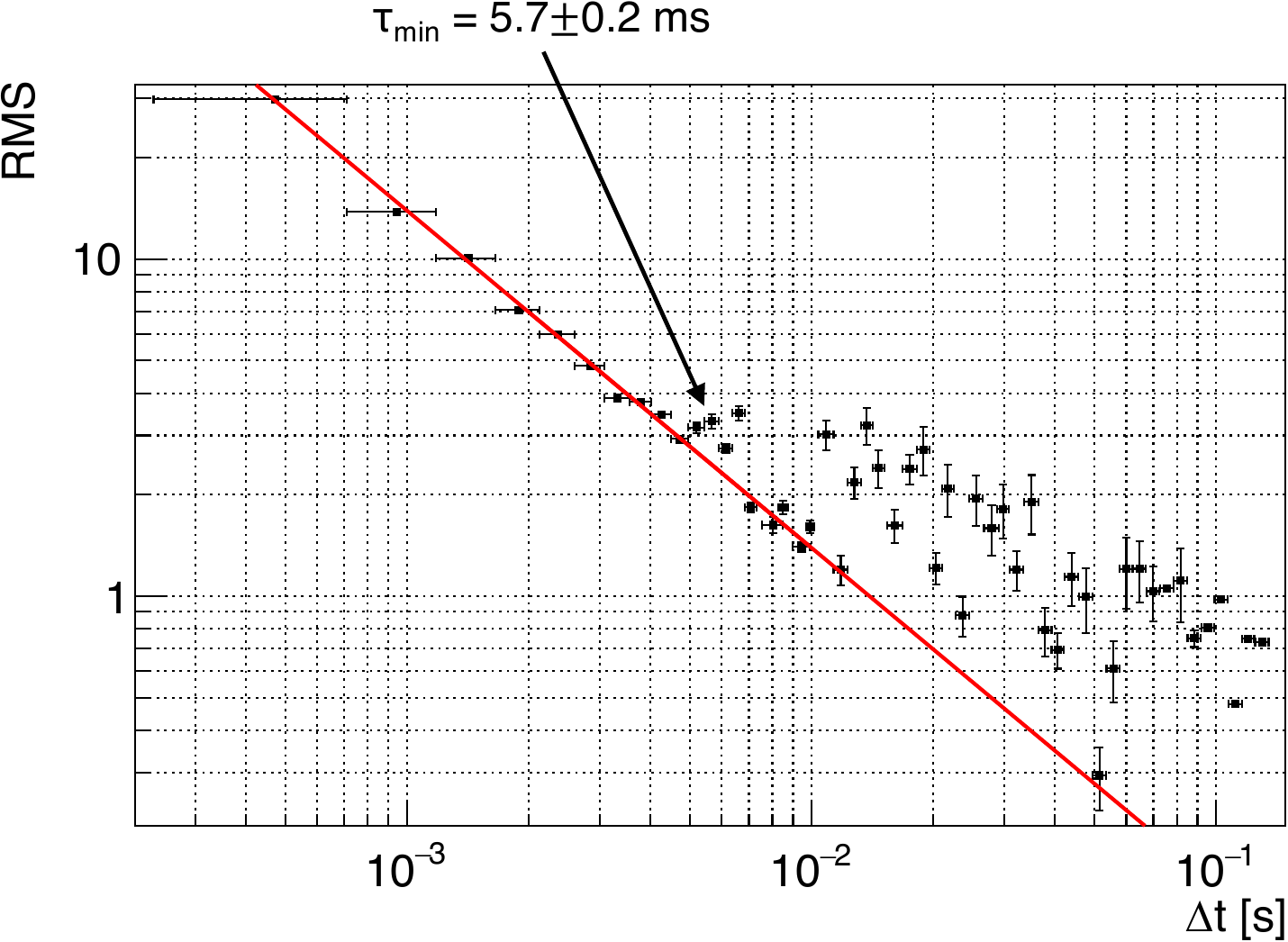}
    \caption{Power spectrum of the root-mean-square (RMS) amplitude of the FREGATE light curve as a function of the variability timescale $\Delta t$ for GRB~050709 at $t_{\rm 0}$ + 0.0--0.5\,s in the energy range of 6--400\,keV. 
    The Poisson noise is represented as $\Delta t^{-1}$ in the power spectrum (red line). The minimum variability timescale ${\tau}_{\mathrm{min}}$ can be derived from the emission of this GRB that exceeds the Poisson-noise component at a confidence level exceeding 3$\sigma$. 
    }
    \label{fig:cmvt}
\end{figure}

\begin{figure}
    \centering
    \includegraphics[width=0.55\textwidth]{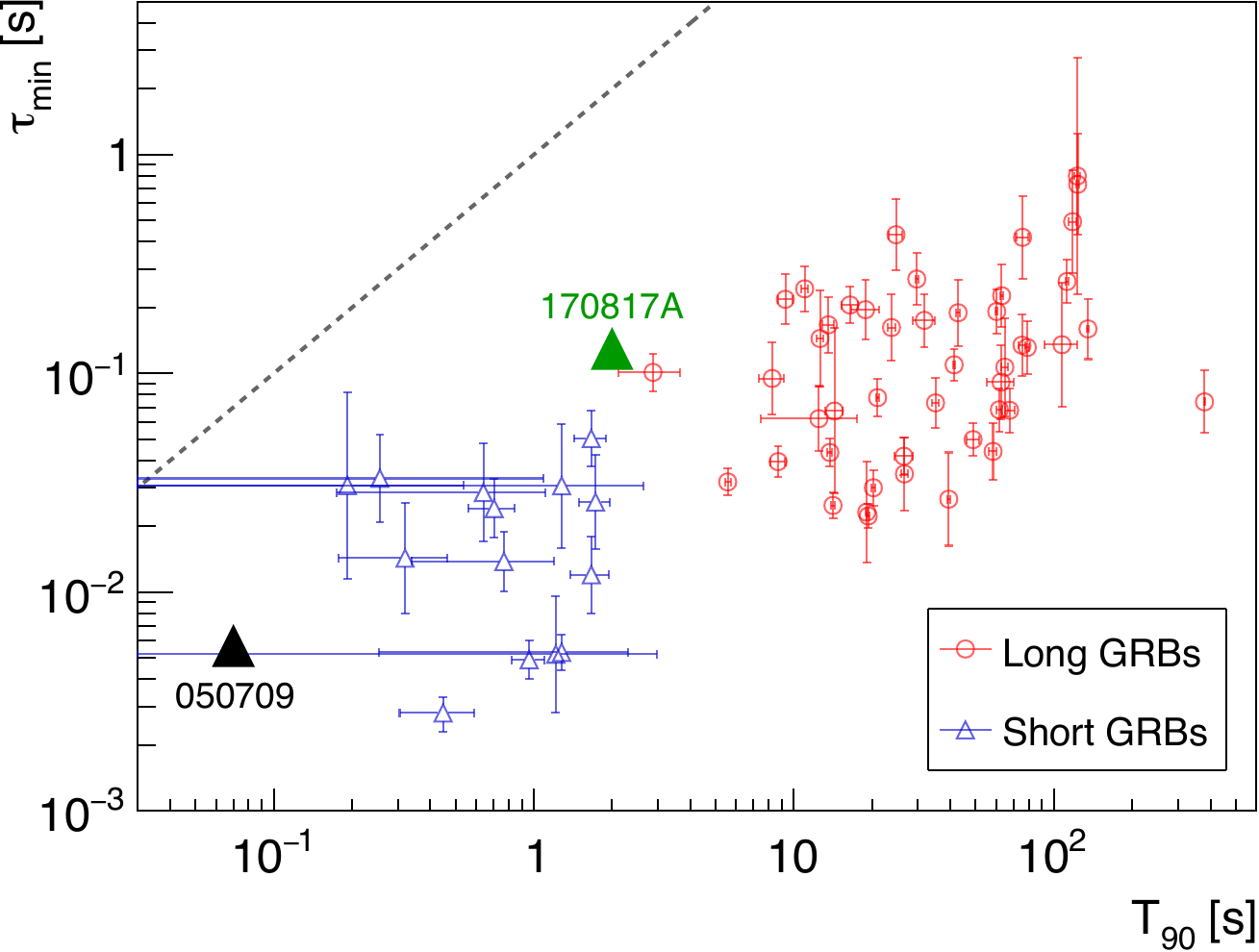}
    \caption{The minimum variability timescale vs. the GRB duration $T_{90}$, for a sample of long and short GRBs. GRB~050709 and GRB~170817A are represented as black and green triangles, respectively. The red circle and blue triangle represent long and short GRBs, respectively, taken from \cite{Maclachlan+2013}.  The minimum variability timescale is calculated using a time interval during the prompt emission (e.g., HS+ST). The dashed black line denotes where values are equal. Note that $T_{90}$ = 70$\pm$10 ms in the 30--400 keV energy band was adopted for GRB~050709.}
    \label{fig:mvt}
\end{figure}

The minimum variability timescale is the time of significant flux variability of emission, which has been considered to give an upper limit on the size of the emitting region and a hint on the nature of the progenitor of the bursts \citep{Schmidt+1978}.
A temporal analysis that determines the power of the differentiated time series of the light curve can be used to estimate the observed variability timescale, $\Delta t$, by subtracting the photon counts in one time bin from the adjacent bin \citep{Nemiroff+1997}.
The obtained power spectrum is shown in Fig. \ref{fig:cmvt} for GRB~050709 using the FREGATE light curve in the energy range of 8--400 keV during the time interval of the short bright emission ($t_{\rm 0}$ + 0.0--0.5 s).
The Poisson noise due to X-ray photon statistics, which can be represented as the power-law function of $\Delta t^{-1}$ in the power spectrum, should be taken into account when estimating the variability timescale from a GRB emission. Thus, we calculated the variability timescale from the emission of this GRB that exceeds the Poisson-noise component with a confidence level greater than 3$\sigma$.  
For GRB~050709, the minimum variability timescale of the short bright emission at $t_{\rm 0}$ + 0.0--0.5 s in the energy range of 8--400 keV is estimated as 5.7$\pm$0.2 ms.
For GRB~170817A observed by \textit{Fermi}/GBM, the minimum variability timescale in the 10--1000 keV energy range was calculated as 
${\tau}_{\mathrm{min}}$ = $0.18{\pm}0.12$\,s with our method, which is consistent with ${\tau}_{\mathrm{min}}$ = 0.13$\pm$0.06\,s estimated by the previous work \citep{Goldstein+2017} using the structure function estimator \citep{2014ApJ...787...90G}.
Fig. \ref{fig:mvt} shows the observed minimum variability timescale ${\tau}_{\mathrm{min}}$ versus the GRB duration $T_{90}$ of GRB 050709, GRB 170817A, and other GRBs. 
This result shows that GRB~050709 belongs to the group of SGRBs 
with the shortest variability timescale and the shortest duration. 
Note that the minimum variability timescales of the hard spike and the soft tail are estimated as ${\tau}^{\mathrm{HS}}_{\mathrm{min}}$ = 2.9$\pm$0.2\,ms and ${\tau}^{\mathrm{ST}}_{\mathrm{min}}$ = 50$\pm$5\,ms, respectively.
In the next section, the estimated minimum variability timescale of the soft tail is used to constrain the emission radius.

\section{Origin of the soft tail}
\label{sec:orig_ST}
Previous studies showed that the electromagnetic emissions of the prompt emission and the afterglow of GRB~170817A  can be explained by
a structured/off-axis relativistic jet \citep{2017MNRAS.471.1652L, 2017Natur.551...71T, 2017ApJ...848L..20M, Alexander+2018, 2017ApJ...848L..25H, 2018MNRAS.478.4128G, 2018ApJ...858L..15D, 2018MNRAS.481.2711G, 2018A&A...613L...1D, 2018PhRvL.120x1103L, 2018NatAs...2..751L, 2018ApJ...856L..18M, Mooley+2018, Troja+2018a,   2019ApJ...870L..15L, 2019ApJ...880L..23W, 2019ApJ...883L...1F,2019ApJ...886L..17H,2020ApJ...896..166R, 2020MNRAS.498.5643T,  2021MNRAS.501.5746T, 2022MNRAS.510.1902T, 2022ApJ...927L..17H}, 
and a jet and cocoon model \citep{Kasliwal+2017, 2017ApJ...834...28N, 2018MNRAS.479..588G, 2019ApJ...887L..16K, 2023MNRAS.520.1111H,2023MNRAS.524.4841H}. 
In particular, observations show that at the prompt emission phase, there is the soft tail emission with a weak thermal component in GRB~170817A\citep{Goldstein+2017}. 
The emission mechanism of this soft tail from GRB~170817A could arise from the photosphere of the jet or the photosphere of the cocoon \citep{Kasliwal+2017,2017ApJ...848L...6L, 2018MNRAS.479..588G,Ioka+2019,2023MNRAS.524.4841H}.

\subsection{Cocoon model}
In the following, we explore the origin of the the soft-tail emission from GRB~050709, and show that it can be explained by cocoon.
First, we assume that the spectrum of the soft tail was observed as a DISKPBB model with $kT_\mathrm{in} \sim 49$\,keV in Table \ref{tab:spec_prompt}.
The isotropic luminosity of the soft tail 
is $L_{\rm iso}^{\rm ST} \sim 6{\times}10^{49}$\,\si{erg.s^{-1}}; using $L=4{\pi}{D_L}^2F$, where $D_L$ is the luminosity distance and $F$ is the energy flux.
$R_{\rm c}$ is the lateral width of the cocoon, immediately after the breakout.
The emitting surface is  $\sim \pi R_{\rm c}^2$, and considering the conical shape of the cocoon, one can roughly find that $R_{\rm c}$ is comparable to the cocoon's photospheric radius \citep{2023MNRAS.524.4841H}
Thus, the cocoon radius $R_{\rm c}$ at the breakout time can be calculated by the following equation:
\begin{equation}
    R_{\rm c} = \left[\frac{{\Gamma}_{\rm c}^2 L_{\rm iso,c}}{4{\pi}{\sigma}\{T_{\rm obs,c}(1+z)\}^4} \left(\frac{\Omega_{\rm EM}}{\Omega}\right)\right]^{\frac{1}{2}}
        = \left[\frac{{\Gamma}_{\rm c}^2 L_{\rm iso,c}}{4{\pi}{\sigma}\{T_{\rm obs,c}(1+z)\}^4} \left(\frac{\theta_{\rm EM}}{\theta_{\rm c}}\right)^2\right]^{\frac{1}{2}} ,
    \label{eq:Rc1}
\end{equation}
where ${\Gamma}_{\rm c}$ is the Lorentz factor of the cocoon, $L_{\rm iso,c}$ is the isotropic equivalent luminosity of the cocoon, $T_{\rm obs,c}$ is the observed cocoon temperature, $\sigma$ is the Stefan–Boltzmann constant, $\Omega_{\rm EM}$ is the solid angle of the electromagnetic emission, $\Omega$ is the solid angle corresponding to the physical size of the cocoon with as $\theta_{\rm c}$ its opening angle, and $\theta_{\rm EM}$ is the opening angle of the electromagnetic emission.
In addition, the opening angle of the emission is $\theta_{\rm EM} \sim {\Gamma}_{\rm c}^{-1}$. 
Thus, Equation \ref{eq:Rc1} becomes
\begin{equation}
    R_{\rm c} =  \left[\frac{L_{\rm iso,c}}{4{\pi}{\sigma}\left\{T_{\rm obs,c}(1+z)\right\}^4} \left(\frac{1}{\theta_{\rm c}}\right)^2\right]^{\frac{1}{2}}
            \sim 2{\times}10^9\,\mathrm{cm} \left(\frac{{\theta}_{\rm c}}{25^{\circ}}\right)^{-1} \left(\frac{kT_{\rm obs,c}}{49\,\mathrm{keV}}\right)^{-2} \left(\frac{L_{\rm iso,c}}{6\times10^{49}\,\mathrm{erg\,s^{-1}}}\right)^{\frac{1}{2}}
    \label{eq:Rc2}
\end{equation}
According to \cite{2023MNRAS.520.1111H}, the typical value for the opening angle of the cocoon is ${\theta}_{\rm c} \sim 25^{\circ}$. 
The maximum cocoon radius $R_{\rm c,max}$ obtained from temporal variability $\tau_{\rm min}^{\rm ST}$ is as follows \citep{Piran+2005}:
\begin{equation}
    R_{\rm c,max} = 4c \frac{\tau_{\rm min}^{\rm ST}}{1+z} {\Gamma}_{\rm c}^2 
            \sim 2\times10^{10}\,\mathrm{cm} \left(\frac{\tau_{\rm min}^{\rm ST}}{50\,\mathrm{ms}} \right) 
            \left(\frac{{\Gamma}_{\rm c}}{2}\right)^2 
            \left(1+z \right)^{-1},
    \label{eq:Rc_max}
\end{equation}
%Here, ${\Gamma}_{c,max}$ is based on the value of ${\Gamma}_{inf}$ in \citet{2023MNRAS.520.1111H} and \citet{2023MNRAS.524.4841H}.
where $c$ is the speed of light.
The typical value for the Lorentz factor of the cocoon ${\Gamma}_{\rm c}$ at the breakout time is a few (see \cite{2023MNRAS.520.1111H, 2023MNRAS.524.4841H}), and we adopt ${\Gamma}_{\rm c} \sim 2$ in this equation. 
The values of $R_{\rm c}$ and $R_{\rm c,max}$ derived from Equations \ref{eq:Rc2} and \ref{eq:Rc_max} confirm $R_{\rm c} < R_{\rm c,max}$.
Assuming a radiation dominant pressure,  the pressure of the cocoon can be calculated by the following equation (assuming an adiabatic index of 4/3):
\begin{equation}
    P_{\rm c} = \frac{1}{4{{\Gamma}_{\rm c}}^2-1} \frac{E_{\rm th}}{V_{\rm c}} {\sim} \frac{1}{4{{\Gamma}_{\rm c}}^2} \frac{E_{\rm th}}{V_{\rm c}}    ,
\end{equation}
where $E_{\rm th}$ is the thermal energy of the cocoon and $V_{c}$ is its volume  (for simplicity ${\Gamma}_{\rm c}\gg1$ here).
Here, the energy density is $E_{\rm th}/V_{\rm c}=4{\sigma}T_{\rm c}^4\Gamma_{\rm c}^2/c$ and $T_{\rm c}={\Gamma_{\rm c}}^{-1}T_{\rm obs,c}$ is the comoving cocoon temperature, giving
\begin{equation}
    P_{\rm c} = \frac{\sigma}{c} \left[\frac{T_{\rm obs,c}(1+z)}{\Gamma_{\rm c}}\right]^4
     \sim 2{\times}10^{19}\,\mathrm{erg\,cm^{-3}} \left(\frac{\Gamma_{\rm c}}{2}\right)^{-4} \left(\frac{kT_{\rm obs,c}}{49\,\mathrm{keV}}\right)^4  
\end{equation}
It is worth noting that the above value of the cocoon pressure is quite consistent with the cocoon pressure as in 2D relativistic hydrodynamical simulation of jet propagation in the expanding ejecta of BNS mergers, as found in \citet{Hamidani+2021}.
\citet{Hamidani+2021} simulated the narrow jet case with the initial opening angle of the jet $\theta_0=6.8^{\circ}$ and $E_{\rm th}\sim 4 \times 10^{46} \mathrm{erg}$, and the wide jet case with $\theta_0=18.0^{\circ}$ and $E_{\rm th}\sim 1 \times 10^{48} \mathrm{erg}$ (see their table 1 and their definition of $\theta_0$)\footnote{Considering the beaming factor of 4$\pi$/$\Omega_{\rm EM} = 1/(1-\cos \theta_{\rm EM})$ with $\theta_{\rm EM}$ = 10--20$^\circ$, we find that $E_{\rm th}/(1-\cos \theta_{\rm EM})$ is of the order of 10$^{49} \mathrm{erg}$ ($\sim E_{\rm iso}^{\rm ST}$).}, showing $\sim$0.2 s after the jet launch the cocoon pressures $P_{\rm c} \sim$ 2$\times$10$^{19}$ erg cm$^{-3}$ and 1$\times$10$^{20}$ erg cm$^{-3}$ and the cocoon radii $R_{\rm c} \sim$ 4$\times$10$^{9}$ cm and 2$\times$10$^{9}$ cm for the narrow- and wide-jet cases, respectively. 
While the estimated cocoon radius of GRB~050709 and the simulated cocoon radii for the narrow- and wide-jet cases are within the same order of magnitude, the estimated cocoon pressure of GRB~050709 is consistent with the narrow-jet case, hence
the narrow-jet scenario might be favored for this GRB in several aspects.

\subsection{Comparison with other GRBs}
%\ma{Here, the difference between the observed BB and DISKPBB temperatures of the soft-tail emission in the 2--400 keV energy range is caused by the spectral shape in the Rayleigh-Jeans range. It should be noted that the {\it HETE-2} observed soft X-ray emission above 2 keV which is lower than the {\it Fermi}-GBM covering X-rays above 8 keV. When we mimicking the {\it Fermi}-GBM observation by limiting the energy range of the {\it HETE-2} above 8 keV, the difference of the BB and DISKPBB models becomes smaller than the case of full energy coverage. Furthermore, considering the fact that the lower-energy observable threshold of the {\it Fermi}-GBM is near }

Table \ref{tab:comparision} shows a comparison between three SGRBs (all having a soft tail component and evidence of kilonova emission in common): GRBs 170817A, 150101B, and 050709. Here, the spectral models for GRBs~170817, 150101B, and 050709 are represented by BB \citep{Goldstein+2017}, BB \citep{Burns+2018}, and DISKPBB (this work), respectively. 
In addition to the observed properties, the temperature and the pressure of these SGRBs are shown, including newly found results here for GRB~050709.
Comparison shows that GRB~050709 has a relatively higher temperature and a much higher pressure compared to the other two SGRBs.
The high temperature and pressure of the cocoon for GRB~050709 can be caused by contribution from the shocked jet part of the \textit{relativistic} cocoon, rather than the shocked ejecta part of the \textit{non-relativistic} cocoon.
This suggests that GRB~050709 was observed with a line-of-sight closer to the jet axis, compared to the other SGRBs, while for GRB~170817A the viewing angle was estimated as $\sim 20^\circ$--$30^\circ$ (\cite{Troja+2018a}, \cite{Mooley+2018}, \cite{Abbott+2017a}).

Another reason for the high temperature of the cocoon in GRB~050709 may be due to the different energy coverage between {\it HETE-2} (2--400 keV) and {\it Fermi}-GBM (8 keV -- 40 MeV; \cite{2009ApJ...702..791M}). The observed spectra of the soft-tail emission for GRBs~150101B and 170817A were well fitted by BB \citep{Goldstein+2017,Burns+2018}. It should be noted that  the lower detectable energy threshold of the {\it Fermi}-GBM ($\sim$8 keV) is close to the observed BB temperatures ($\sim$6 keV for GRB~150101B and $\sim$10 keV for GRB~1708017A; \cite{Burns+2018,Goldstein+2017}), in which case the spectral slope in the Rayleigh-Jeans range cannot be accurately determined and the BB model may sufficiently fit the observed spectrum. For GRB~050709, the BB model yields a temperature ($\sim$17 keV) similar to that of the other GRBs. However, a broader energy coverage, with the lower detectable energy threshold of the {\it HETE-2} ($\sim$2 keV), favors a modified BB model such as DISKPBB with $kT_{\rm in} \sim 49$ keV over BB. 
Here, we mimicked the {\it Fermi}-GBM observations by constraining the energy band of {\it HETE-2} above 8 keV (See the results in the 8--400 keV band of Table \ref{tab:spec_prompt}).  The results in the 8--400 keV energy band show a smaller difference in the PBstat values between the BB and DISKPBB models compared to those in the 2--400 keV energy band: $\Delta$PGstat values between BB and DISKPBB are  approximately 8 and 3 in the 2--400 keV and 8--400 keV energy bands, respectively. This suggests that observations with a higher energy threshold make it challenging to distinguish between the BB and modified BB spectra. 
Note that if the representative model for GRB~150101B and 170817A is a modified BB with observations covering a wider energy range, the observed temperatures could be higher than that of a pure BB, potentially resulting in different cocoon radius and cocoon pressure: smaller $R_{\rm c}$ and higher $P_{\rm c}$.

For GRB 150101B, assuming that the soft-tail emission is represented by a BB, the cocoon radius is calculated to be $R_{\rm c}>R_{\rm c,max}$, which contradicts the cocoon scenario and may suggest that the soft-tail emission arises from the photoshere of the ultra-relativistic jet \citep{2000ApJ...530..292M}, 
rather than from the photoshere of the cocoon. In the case of the relativistic jet, the photospheric radius is estimated as
$R_{\rm ph} \sim 10^{10}$ $\left(\Gamma_{\rm j}/100 \right)^{-3} \left(L_{\rm iso,j}/10^{51.8} \mathrm{erg\,s^{-1}} \right)$
cm, where $\Gamma_{\rm j}$ is the bulk Lorentz factor of a jet and $L_{\rm iso,j}$ is the total jet luminosity assumed to be approximately 10 times $L_{\rm iso}$ \citep{Burns+2018},
while the maximum jet emission radius is  $R_{\rm j,max}$ $\sim$ 10$^{13}$ cm $\left(\tau_{\rm min}^{\rm ST}/16\,{\rm ms} \right) \left(\Gamma_{\rm j}/100 \right)^2$. Furthermore, the photospheric temperature and luminosity can be estimated as 
$kT_{\rm ph}$ $\sim$ 6 $\left(\Gamma_{\rm j}/100 \right)^{8/3}$ 
$\left(R_{\rm 0}/10^{6.4}\,{\rm cm} \right)^{1/6} 
\left(L_{\rm iso,j}/10^{51.8} \mathrm{erg\,s^{-1}} \right)^{-5/12} \mathrm{keV}$ 
and  $L_{\rm iso,ph}$ $\sim$ 10$^{49.2}$ $\left(\Gamma_{\rm j}/100 \right)^{8/3}$ 
%$\left(R_{\rm 0}/10^{6.4}\,{\rm cm} \right)^{2/3} \left(L_{\rm iso,j}/10^{51.8} \mathrm{erg\,s^{-1}} \right)^{1/3} \mathrm{erg\,s^{-1}}$, respectively, where $R_0$ is the innermost stable circular orbit for a 2.8 M$_\solar$ black hole corresponding to the total mass of GW~170817 \citep{Abbott+2017a}, and these estimates are consistent with the observed values.
$\left(R_{\rm 0}/10^{6.4}\,\mathrm{cm} \right)^{2/3} \left(L_{\rm iso,j}/10^{51.8} \mathrm{erg\,s^{-1}} \right)^{1/3} \mathrm{erg\,s^{-1}}$, respectively, where $R_0$ is the innermost stable circular orbit for a 2.8 $M_{\odot}$ black hole corresponding to the total mass of GW~170817 \citep{Abbott+2017a}, and these estimates are consistent with the observed values.
When considering the photospheric jet scenario for GRB~050709 with an assumed $\Gamma_{\rm j} \sim 270$, the photospheric temperature and luminosity can be estimated as 
$kT_{\rm ph}$ $\sim$ 50 $\left(\Gamma_{\rm j}/270 \right)^{8/3}$ 
$\left(R_{\rm 0}/10^{6.4}\,{\rm cm} \right)^{1/6} 
\left(L_{\rm iso,j}/10^{51.3} \mathrm{erg\,s^{-1}} \right)^{-5/12} \mathrm{keV}$ 
and  $L_{\rm iso,ph}$ $\sim$ 10$^{49.8}$ $\left(\Gamma_{\rm j}/270 \right)^{8/3}$ 
$\left(R_{\rm 0}/10^{6.4}\,{\rm cm} \right)^{2/3} \left(L_{\rm iso,j}/10^{51.3} \mathrm{erg\,s^{-1}} \right)^{1/3} \mathrm{erg\,s^{-1}}$, respectively. Those estimated values are also consistent with the observed values, suggesting that the cocoon scenario for GRB~050709 is one of the plausible scenarios, and other scenarios, such as the photospheric jet, cannot be ruled out. Discrimination and constraints on the models require a determination of physical parameters such as the bulk Lorentz factor. 
While afterglow modeling from follow-up multi-wavelength observations provides meaning constraints on the bulk Lorentz factor, the scarcity of data points in the follow-up observations makes it challenging for this GRB.

\begin{table*}
    \tbl{Comparison of the physical parameters of the soft-tail emission as estimated in Section 3, for different SGRBs with evidence of kilonova emission. 
    }{
    \resizebox{\textwidth}{!}{
    \begin{tabular}{llll} \hline \hline
                                                    & 170817A               
                                                    & 150101B          
                                                    & 050709 (This work)
                                                    \\ \hline
        Observed temperature $kT_{\rm obs}$ [keV] & $10.3{\pm}1.5$ \citep{Goldstein+2017} 
                                                    & $6.0{\pm}0.6$ \citep{Burns+2018}     
                                                    & $49^{+15}_{-19}$  
                                                    \\
        %\ogino{Viewing angle $\theta_v$}            & \ogino{$14^{\circ}$--$28^{\circ}$ \citep{Mooley+2018}}
        %                                            & \ogino{$8^{\circ}$--$20^{\circ}$ \citep{Troja+2018b}} 
        %                                            & \ogino{$\sim10^{\circ}$?} \footnotemark                                   
        %                                            \\ 
        %\ogino{Doppler factor $\delta$ $({\Gamma}=5)$}  & 2.3 ($\theta_v=21^{\circ}$)                   
        %                                              & 4.1 ($\theta_v=14^{\circ}$)
        %                                             & 5.7 ($\theta_v=10^{\circ}$)                                          
        %                                            \\
        Isotropic luminosity $L_{\rm iso}^{\rm ST}$ [\si{erg.s^{-1}}]  & $10^{46.1}$      
                                                              & $10^{49.1}$   
                                                              & $10^{49.8}$
                                                              \\
        %\ogino{On-axis cocoon luminosity $L_{on,c}=L_{obs,c}(0^{\circ})$ [\si{erg.s^{-1}}]} & $3\times10^{47}$   
        %                                                                                    & $4\times10^{51}$    
        %                                                                                    & $6\times10^{52}$                                         
        %                                                                            \\
        %Variability timescale (soft tail) \ogino{$\tau_{min}^{ST}$} [ms]     & $220{\pm}128$         
        %                                                                     & $16{\pm}5$         
        %                                                                     & $50{\pm}16$                                                  
        %                                                                     \\ 
        Cocoon radius $R_{\rm c}$ [cm] & $6\times10^{8}
                                                   \Bigl(\frac{kT_{\rm obs}}{10.3\,\mathrm{keV}}\Bigr)^{-2}
                                                   \Bigl(\frac{L_{\rm iso}}{10^{46.1}}\Bigr)^{\frac{1}{2}}$
                                                 & $4\times10^{10}
                                                   \Bigl(\frac{kT_{\rm obs}}{6.0\,\mathrm{keV}}\Bigr)^{-2}
                                                   \Bigl(\frac{L_{\rm iso}}{10^{49.1}}\Bigr)^{\frac{1}{2}}$ 
                                                 & $2\times10^{9}
                                                   \Bigl(\frac{kT_{\rm obs}}{49\,\mathrm{keV}}\Bigr)^{-2}
                                                   \Bigl(\frac{L_{\rm iso}}{10^{49.8}}\Bigr)^{\frac{1}{2}}$  
                                                \\
        Maximum Cocoon radius $R_{\rm c,max}$ [cm] & $1\times10^{11} 
                                                                 \Bigl(\frac{\tau_{\rm min}^{\rm ST}}{225\,\mathrm{ms}}\Bigr)
                                                                 \Bigl(\frac{{\Gamma}_{\rm c}}{2}\Bigr)^2$
                                                               & $8\times10^{9}   
                                                                 \Bigl(\frac{\tau_{\rm min}^{\rm ST}}{16\,\mathrm{ms}}\Bigr)
                                                                 \Bigl(\frac{{\Gamma}_{\rm c}}{2}\Bigr)^2$
                                                               & $2\times10^{10} 
                                                                 \Bigl(\frac{\tau_{\rm min}^{\rm ST}}{50\,\mathrm{ms}}\Bigr)
                                                                 \Bigl(\frac{{\Gamma}_{\rm c}}{2}\Bigr)^2$  
                                                               \\
        Cocoon pressure $P_{\rm c}$ [\si{erg.cm^{-3}}] & $3{\times}10^{16} \Bigl(\frac{{\Gamma_{\rm c}}}{2}\Bigr)^{-4} 
                                                                   \Bigl(\frac{kT_{\rm obs}}{10.3\,\mathrm{keV}}\Bigr)^{4}$
                                                                 & $5{\times}10^{15} \Bigl(\frac{{\Gamma_{\rm c}}}{2}\Bigr)^{-4} 
                                                                   \Bigl(\frac{kT_{\rm obs}}{6.0\,\mathrm{keV}}\Bigr)^{4}$
                                                                 & $2\times10^{19} \Bigl(\frac{{\Gamma_{\rm c}}}{2}\Bigr)^{-4} 
                                                                   \Bigl(\frac{kT_{\rm obs}}{49\,\mathrm{keV}}\Bigr)^{4}$
                                                                \\ \hline
    \end{tabular}}}
    \label{tab:comparision}
    \leavevmode \\{
    Note: The redshift values of GRBs~170817A, 150101B and 050709 are 0.00968 \citep{Coulter+2017}, 0.134 \citep{Fong+2016}, and 0.16 \citep{Fox+2005}, respectively. The thermal spectra for GRBs~170817, 150101B, and 050709 are represented as BB \citep{Goldstein+2017}, BB \citep{Burns+2018}, and DISKPBB, respectively.
    }
\end{table*}

%\footnotetext{\ogino{The Lorenz factor of GRB~150101B was set $\Gamma_c=5$ to satisfy $R_c{\lesssim}R_{c,max}$. This is a larger value than the other two GRBs, but it satisfies definition of cocoon at \citet{2023MNRAS.524.4841H}.}}

\section{Origin of the soft extended emission}
\label{sec:orig_EE}
\begin{figure*}
    \centering
    \includegraphics[width=0.6\textwidth]{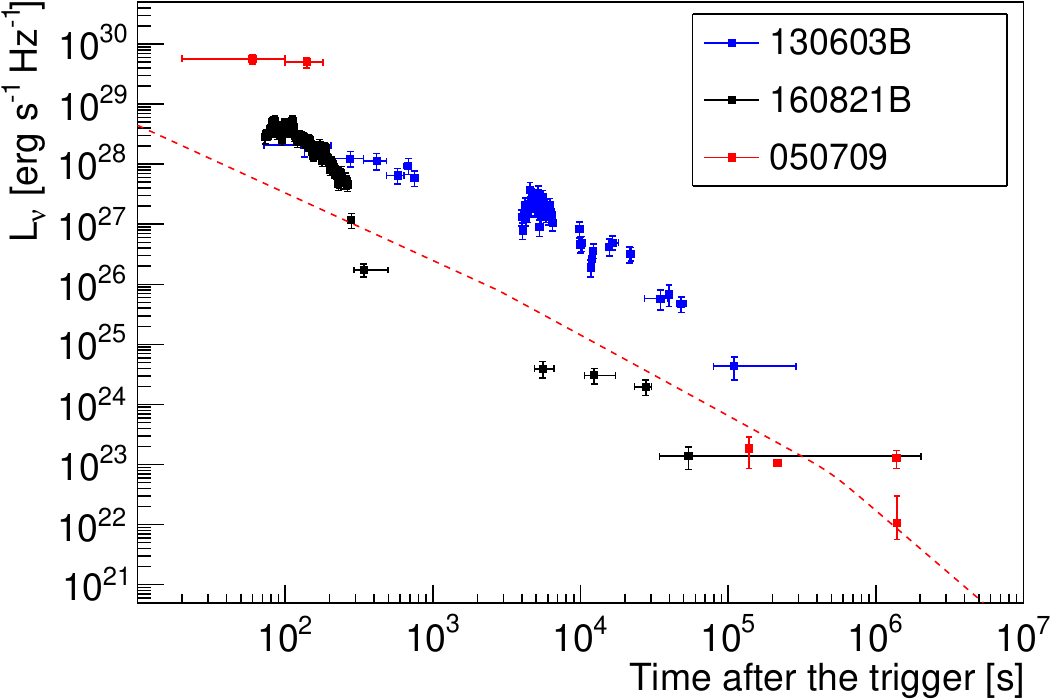}
    \caption{The observed luminosity density lightcurves at 5 keV for SGRBs~050709, 130603B and 160821B that have kilonova association. The red dashed line represents the afterglow model for GRB~050709 with $\varepsilon_{\rm e}$ = 0.1, $\varepsilon_{B}$ = 0.05, and $n_{\rm ISM}$ = 10$^{-2}$ cm$^{-3}$.}
    \label{fig:flux_density}
\end{figure*}

We compared the soft X-ray light curve of GRB~050709 with those of SGRBs with the extended emission and kilonova, GRBs~130603B \citep{2013Natur.500..547T} and 160821B \citep{2017ApJ...843L..34K}, as shown in Fig. \ref{fig:flux_density}.
Although the number of data points in the X-ray light curve for GRB~050709 is sparse, the soft X-ray emission during the first $\sim$100 seconds is much higher than that of GRBs~130603B and 160821B by one to two orders of magnitude. Note that redshift values are $z=0.16$,  0.36, and 0.16 for GRBs~050709, 130603B and 160821B \citep{2016GCN.19846....1L, 2014A&A...563A..62D}.
 Several short GRB spectra (e.g., GRB~160821B) show a substantial spectral softening in the early decay phase during the first a few hundred seconds \citep{Kagawa+2015,2019ApJ...877..147K}. 
 %In contrast, as shown in Table \ref{tab:EE} the soft extended emission of GRB~050709 exhibits no significant spectral softening.  This indicates that the soft extended emission may arise from different mechanisms:
 As shown in Table \ref{tab:EE} the soft extended emission of GRB~050709 may exhibit no significant spectral softening.  This indicates that the soft extended emission may arise from different mechanisms:
 One can predict the dissipation of a Blandford–Znajek outflow \citep{2011MNRAS.417.2161B, 2014ApJ...796...13N} and the spin-down of a highly magnetized rapidly rotating neutron star \citep{2008MNRAS.385.1455M,2012MNRAS.419.1537B}.  Those models anticipate that the soft extended emission originates from a mildly relativistic jet component different from the prompt hard-spike jet.

Another scenario for explaining the soft extended emission might be afterglow, which allows reconciliation with the late emission at $t_0$ + 10$^5$--10$^6$ s using fiducial parameters (with $\varepsilon_{\rm e}$ = 0.1, $\varepsilon_{B}$ = 0.05, and $n_{\rm ISM}$ = 10$^{-2}$ cm$^{-3}$, where $\varepsilon_{\rm e}$ and $\varepsilon_{B}$ represent the fractions of energy in the electrons and the magnetic fields, respectively, and $n_{\rm ISM}$ is the density of the interstellar medium). However,  it fails to explain the soft extended emission at $t_0$ + $\sim$100 s: the observed soft extended emission clearly exceeds the afterglow model, as illustrated in Fig. \ref{fig:flux_density}. Such a similar feature is also observed in GRB~160821B.
It should be noted that more complex afterglow models are still possible (e.g., \cite{2019ApJ...883...48L}).

\section{Conclusion}
\label{sec:conclusion}
We report the possible evidence of soft X-ray emission right after the prompt hard-spike emission from GRB~050709, which may be interpreted as cocoon emission. The cocoon emission is tightly coupled to the jet physics of GRBs.
Hence, assuming radiation dominant cocoon
we estimated the properties of the cocoon of GRB~050709 at the breakout time such as the cocoon radius as $\sim$10$^{9}$ cm and the cocoon pressure as $\sim$10$^{19}$ erg cm$^{-3}$.
These results are consistent with the cocoon model with injected thermal energy of $\sim$$10^{46} \mathrm{erg}$ given by the previous numerical studies in the narrow-jet scenario \citep{Hamidani+2021}. 
We also compared the results of GRB~050709 to those of other short GRBs, specifically GRBs 170817A and 150101B, which exhibited similar light curves (i.e., hard-spike and soft-tail components). Our analysis indicates that the cocoon pressure and temperature for GRB~050709 are relatively higher than those observed in the other two GRBs. This difference may suggest that GRB~050709 had an on-axis jet compared to the other two GRBs.
The viewing angle provides crucial information to understand GRB brightness, intrinsic rate, jet structure \citep{2001ApJ...554L.163I,2002ApJ...571L..31Y, 2004ApJ...607L.103Y,2017ApJ...851L..32E,2018PTEP.2018d3E02I,2018ApJ...869L...4E,2019MNRAS.486.1563M,Ioka+2019,2019ApJ...884...71F,2021MNRAS.501.5746T} and muti-wavelength afterglow lightcurves (e.g., \cite{Troja+2018a,Troja+2018b,2019ApJ...883L...1F,2019MNRAS.489.2104T,2020ApJ...899..105L,2022ApJ...940..189F}), along with details on kilonova characteristics because the viewing angle may also contribute to explaining the observed complex behavior of the kilonova emission from GW~170817 (e.g., \cite{2017PASJ...69..102T, 2020ApJ...889..171K}).

The soft tail emission, as discussed in the paper, was predominantly observed below $\sim$10 keV, as seen in Fig. \ref{fig:lc_prompt}. If this holds for other SGRBs, current observatories such as \textit{Fermi}-GBM and \textit{Swift}-BAT face challenges in effectively covering this soft X-ray band. The future gamma-ray burst mission, HiZ-GUNDAM \citep{2020SPIE11444E..2ZY}, scheduled for the 2030s, is expected to advance capabilities in detecting electromagnetic emission from SGRBs: the wide-field X-ray monitors will be able to detect soft X-ray emission in the prompt and extended emission phases, and the near infrared telescope will conduct the follow-up observations of early afterglow or kilonova emission, which makes this mission very unique.

%%%%%%%%%%%%%%%%%%%%%%%%%%%%%%%%%%%%%%%

\begin{ack}
   This study was supported by JSPS KAKENHI Grant Numbers JP22J12717 (N.O.),
   JP17H06362 (M.A.), JP23H04898 (D.Y. and M.A.), JP23H04895 (T.S.), the CHOZEN Project of Kanazawa University (D.Y., M.A., and T.S.), 
   and the JSPS Leading Initiative for Excellent Young Researchers Program (M.A.).
\end{ack}

%%%
% See the manual for the detail.
%%%
%\bibliographystyle{plainnat}
%\bibliographystyle{plainnat}
\bibliographystyle{aasjournal}
\bibliography{main_body}

%\begin{thebibliography}{}
%% Journals(e.g. A\&A,ApJ,AJ,NMRAS,PASP ...)
%% Authors, Year, Journal, Vol#, Page#
%% Journal Title Abbreviation >> http://www.asj.or.jp/pasj/Jabb.html
%\bibitem[Fox et al. (2005)]{Fox+2005}
%  Fox, Derek B et al.\ 2005, Nature, 437, 845--850
%\bibitem[Shirasaki et al. (2003)]{Shirasaki+2003}
%  Shirasaki, Y., et al.\ 2003, Publications of the Astronomical Society of Japan, 55, %1033--1049
%\bibitem[Atteia et al. (1995)]{Atteia+1995}
%  Atteia, J-L., et al.\ 1995, Astrophysics and Space Science, 231, 471--474
%\bibitem[Ricker et al. (2001)]{Ricker+2001}
%  Ricker G. R., et al.\ 2003, AIP Conf. Proc., 662, 3--16
%\bibitem[Villasenor et al. (2005)]{Villasenor+2005}
%  Villasenor, J., et al.\ 2005, Nature, 437, 855--858
%\bibitem[Goldstein et al. (2017)]{Goldstein+2017}
%  Goldstein, A., et al.\ 2017, ApJ, 848.2, L14
%\bibitem[Norris \& Bonnell (2006)]{Norris+2006}
%  Norris J. P. and Bonnell J. T.\ 2006, ApJ, 643.1, 266
%\bibitem[Kagawa et al. (2015)]{Kagawa+2015}
%  Kagawa, Y., et al.\ 2015, ApJ, 811.1,  4
%\bibitem[Hamidani et al. (2021)]{Hamidani+2021}
%  Hamidani, H., et al.\ 2021, Monthly Notices of the Royal Astronomical Society, 500.1, %627--642.
%\end{thebibliography}

\end{document}